\documentclass[]{acmart}%review

\AtBeginDocument{%
  \providecommand\BibTeX{{%
    \normalfont B\kern-0.5em{\scshape i\kern-0.25em b}\kern-0.8em\TeX}}}

\setcopyright{acmcopyright}
\copyrightyear{2022}
\acmYear{2022}
\acmDOI{XXXX}

\acmPrice{15.00}
\acmISBN{978-1-4503-XXXX-X/18/06}

\acmConference[Conference acronym 'XX]{Make sure to enter the correct conference title from your rights confirmation email}{June 03--05, 2018}{Woodstock, NY}
\acmBooktitle{Woodstock '18: ACM Symposium on Neural Gaze Detection, June 03--05, 2018, Woodstock, NY}

\usepackage[caption=false,font=normalsize,labelfont=sf,textfont=sf]{subfig}

\usepackage{stfloats}
\usepackage{enumerate}

\usepackage{tabularx, makecell, multirow, balance, booktabs}
\usepackage[flushleft]{threeparttable}
\usepackage[htt]{hyphenat}
\usepackage{seqsplit}

\usepackage{url}
\usepackage[most]{tcolorbox}

\tcbset{textmarker/.style={%
        enhanced,
        parbox=false,boxrule=0mm,boxsep=0mm,arc=0mm,
        outer arc=0mm,left=1.8mm,right=1.8mm,top=7pt,bottom=7pt,
        toptitle=1mm,bottomtitle=1mm,oversize}}

\newtcolorbox{noteBox}{textmarker,
    borderline west={3pt}{0pt}{Gray},
    colback=gray!10!white}

\usepackage{letltxmacro}
\LetLtxMacro\oldttfamily\ttfamily
\DeclareRobustCommand{\ttfamily}{\oldttfamily\csname ttsize\endcsname}
\newcommand{\setttsize}[1]{\def\ttsize{#1}}
\setttsize{\small}% \ttfamily will be \Large

\begin{document}

\title[Cybersecurity Discussions in Stack Overflow]{Cybersecurity Discussions in Stack Overflow: A Developer-Centred Analysis of Engagement and Self-Disclosure Behaviour}

%Characterizing Engagement in Stack Overflow: Analyzing Cybersecurity Discussions And Self-Disclosure Behavior Among Software Developers

%Characterizing Self-Disclosure in Stack Overflow: Analysing Profile Visibility and Cybersecurity Discussions Among Software Developers

%Characterizing Cybersecurity Discussions in Stack Overflow: Analyzing Profile Visibility and Developers' Engagement 

%Characterizing Participation in Stack Overflow: Analyzing Profile Visibility and Engagement in Cybersecurity Discussions.

%Analyzing the Interplay Between Between Profile Visibility and Involvement in Cybersecurity Discussions in StackOVerflow.

%Who Contributes To Cybersecurity? Analyzing Profile Visibility and Cybersecurity Discussions in Stack Overflow.

%%
\author{Nicolás E. Díaz Ferreyra}
\email{nicolas.diaz-ferreyra@tuhh.de}
\orcid{0000-0001-6304-771X}
\affiliation{%
  \institution{Hamburg University of Technology}
  %\city{Hamburg}
  \country{Germany}
}

\author{Melina Vidoni}
\email{melina.vidoni@anu.edu.au}
\orcid{0000-0002-4099-1430}
\affiliation{%
  \institution{Australian National University}
  \country{Australia}
}

\author{Maritta Heisel}
\email{maritta.heisel@uni-due.de}
\affiliation{%
  \institution{University of Duisburg-Essen}
  %\city{Duisburg}
  \country{Germany}
}

\author{Riccardo Scandariato}
\email{riccardo.scandariato@tuhh.de}
\affiliation{%
  \institution{Hamburg University of Technology}
  %\city{Duisburg}
  \country{Germany}
}

\renewcommand{\shortauthors}{Díaz Ferreyra et al.}

%% NICO: EL ABSTRACT ES EL MISMO SOLO QUE LO PARTI EN LOS SEGMENTOS QUE PIDEN. VOS REACOMODALO.
%% Segmentos: Background, Objective, Method, Results, Conclusion)
%%  Colega mio sugirio que cada fragmento del abstract no tenga mas de 2 oraciones, o 3 cortas
\begin{abstract}
Stack Overflow (SO) is a popular platform among developers seeking advice on various software-related topics, including privacy and security. As for many knowledge-sharing websites, the value of SO depends largely on users' engagement, namely their willingness to answer, comment or post technical questions. Still, many of these questions (including cybersecurity-related ones) remain unanswered, putting the site's relevance and reputation into question. Hence, it is important to understand users' participation in privacy and security discussions to promote engagement and foster the exchange of such expertise.
\textbf{Objective}: Based on prior findings on online social networks, this work elaborates on the interplay between users' engagement and their privacy practices in SO. Particularly, it analyses developers' self-disclosure behaviour regarding profile visibility and their involvement in discussions related to privacy and security.
\textbf{Method}: We followed a mixed-methods approach by (i) analysing SO data from 1239 cybersecurity-tagged questions along with 7048 user profiles, and (ii) conducting an anonymous online survey (N=64). 
\textbf{Results}: About 33\% of the questions we retrieved had no answer, whereas more than 50\% had no accepted answer. We observed that \textit{proactive} users tend to disclose significantly less information in their profiles than \textit{reactive} and \textit{unengaged} ones. However, no correlations were found between these engagement categories and privacy-related constructs such as \textit{perceived control} or \textit{general privacy concerns}. \textbf{Implications}: These findings contribute to (i) a better understanding of developers' engagement towards privacy and security topics, and (ii) to shape strategies promoting the exchange of cybersecurity expertise in SO.
\end{abstract}



\maketitle

\section{Introduction}\label{introduction}

The last decade has put privacy in the spotlight of software development, as new legal frameworks emerged to safeguard people's data protection rights and promote responsible engineering practices. One clear example is the EU General Data Protection Regulation (GDPR) \cite{regulation2016regulation} which has introduced strong legal provisions seeking to enforce software companies to comply with a set of privacy principles including transparency, fairness, and informed consent. More recently, as the software industry moves towards the development of Artificial Intelligence (AI) applications, a new regulatory framework is in sight \cite{ai_act2021}, promising to strengthen the protection and governance of personal data in AI systems. In turn, companies and organisations have been urged to adopt privacy-by-design practices to comply with current regulations. Nevertheless, this has also raised questions and concerns among software developers on how to effectively translate these legal provisions and privacy principles into technical solutions \cite{sirur2018we}. %Are We There Yet?

Question-Answer (Q\&A) platforms are a valuable resource for both experienced and junior programmers seeking support in their software development tasks. Stack Overflow (SO) \cite{stackOverflow2021} is among the largest Q\&A platforms in which developers participate in discussions related to performance issues, bugs, and code workarounds \cite{ahmed2017understanding}. Given the increasing importance of cybersecurity in software engineering, a large number of questions regarding privacy, security, and data protection have been posited and addressed by SO users. Particularly, issues related to GDPR compliance, privacy policies, and access-control are some of the most popular privacy-related discussions in SO \cite{tahaei2020understanding,lopez2018sead}. Still, privacy- and security-related topics receive little attention in comparison to others such as data science, big data, and mobile operating systems\footnote{By May 2021, the amount of \textit{security-} and \textit{privacy-}related questions was around 53.000, whereas for \textit{Android} and \textit{iOS} it was over 1.900.000  \cite{stackTags2021}.}. Albeit this suggests a low engagement towards cybersecurity discussions within the SO community, it also reveals an overall tendency among software developers to overlook privacy and security aspects of their code~\cite{senarath2018ease,assal2018motivations,hadar2018privacy}.

\subsection{Motivation}

Developers play a key role in embedding privacy and security principles into the core architecture of information systems \cite{hadar2018privacy}. However, many often fail to create secure software solutions that successfully preserve users' privacy and data protection rights \cite{hadar2018privacy,senarath2018ease}. Over the last years, a growing body of research has leveraged the SO's dataset to identify and characterise cybersecurity trends among software practitioners. Prior work has investigated developers' motivations \cite{lopez2018sead}, knowledge gaps \cite{tahaei2020understanding}, and concerns towards privacy and security \cite{Lopez2019}. However, ``\textit{answer-hungry}'' questions are still a common phenomenon and an ongoing issue within Q\&A websites (i.e., questions remaining unanswered or unresolved) \cite{gao2020technical}. Being SO a community frequented by more than 100 Million developers per month~\cite{stackOverflowUsrs}, users' commitment towards timely and high-quality answers becomes critical for the platform's reputation and success. Former research has sought to understand users' motivations (and amotivations) when it comes to participation in Q\&A forums \cite{yang2014sparrows,chua2015answers,adaji2016towards}. Yet, little effort has been made to characterise users' engagement in cybersecurity discussions in SO. That is, on providing evidence and actionable information about community members participating actively (or not) in such exchanges.

%Although some effort has been put in the identification of 

%Although efforts have been made to understand users' motivations (and amotivations) towards Q\&A participation, 

%it is critical for its members to find answers timely. 

%If we consider that the Q\&As posted within this community are consulted daily by more than 18 million users, it is critical to identify  

%Moreover, that SO is consulted daily by a community of more than 18 million users, it is critical

%essential to understand the factors contributing to or impairing users' participation in cybersecurity questions. Furthermore, as research has 

%Nonetheless, little effort has been made regarding the identification of nuances between users with major cyber-security concerns and those showcasing low levels of privacy and security awareness. That is, on providing evidence and pointers towards practitioners who may (or may not) require further support and guidance in either (i) adopting secure coding practices, or (ii) incorporating privacy principles into their software.

%The interplay between users' privacy concerns, self-disclosure behavior, and engagement has been extensively investigated within the scope of Online Social Networks (ONSs) like Facebook or Twitter \cite{ostendorf2020neglecting,kramer2019mastering,adhikari2018users,dincelli2017can}. 

Individuals' engagement in Online Social Networks (OSNs) like Facebook has been extensively investigated from the perspective of privacy concerns. Such research has analysed the connection between users' self-disclosure decisions (e.g., the amount of private information they reveal inside profiles and posts) and their engagement in these platforms (e.g., number and quality of OSN posts) \cite{kayes2015privacy,choi2018instagram,staddon2012privacy}. Overall, such research has not only contributed to a better understanding of users' privacy concerns and practices but has also paved the road for the development of user-centred technologies. That is, for the elaboration of methods and tools aiming to support and guide users' interaction in OSN environments \cite{seamons2022privacy}. However, to the extent of our knowledge, the role of privacy-related behaviour has not been closely investigated within Q\&A platforms like SO. Particularly, the interplay between developers' self-disclosure practices and their engagement in discussion threads has not been yet explored under the lens of privacy and security benchmarks. 

%Individuals' privacy behavior has been extensively investigated within the scope of Online Social Networks (OSNs) like Facebook or Twitter \cite{ostendorf2020neglecting,kramer2019mastering,adhikari2018users,dincelli2017can}. Prior research has analyzed the connection between users' self-disclosure decisions (e.g., the amount of private information they reveal inside profiles and posts) \cite{ostendorf2020neglecting}, engagement (e.g., number and quality of OSN posts) \cite{kayes2015privacy}, and cybersecurity practices (e.g., their tendency towards stronger password schemas) \cite{dincelli2017can}. Overall, such research has not only contributed to a better understanding of user's privacy concerns and practices, but it has also paved the road for the development of user-centred technologies. That is, for the elaboration of methods and tools aiming to support users' privacy and security practices in OSN environments. However, to the extent of our knowledge, the role of privacy-related behavior has not been closely investigated within Q\&A platforms like SO. Particularly, the interplay between developers' self-disclosure practices and their engagement in cybersecurity discussions has not been yet explored under the lens of privacy and security benchmarks. 

\subsection{Contribution and Research Questions}

SO is a valuable resource for developers seeking advice about multiple aspects of software development. Given the increasing importance of cybersecurity in software engineering, it becomes necessary to foster the engagement among its users towards privacy- and security-related discussions. Hence, this work aims at contributing to ongoing research in SO by investigating the interplay between users' self-disclosure decisions and their engagement in cybersecurity discussions. All in all, the research questions (RQs) this paper seeks to answer are:

%Developer-centred security is an emerging discipline elaborating on user-centred methods and tools for supporting software engineers in their cybersecurity decisions \cite{tahaei2019survey}. For this purpose, such a discipline nourishes from research addressing the factors that may impair (or motivate) developers to adopt privacy- and security-by-design practices. Hence, this work aims at contributing to ongoing research in this area by investigating the self-disclosure decisions of SO users along with their engagement in cybersecurity discussions. All in all, the research questions (RQs) this paper seeks to answer are:

\begin{itemize}
    \item \textbf{RQ1}: \textit{Are users' self-disclosure behaviour associated with their engagement in cybersecurity discussions?} Prior studies in OSNs (in general) and Q\&A platforms (in particular) have shown correlations between users' engagement and self-disclosure practices (e.g., \cite{vargo2018identity,adaji2016towards,kayes2015privacy}). Hence, this RQ aims at zooming into developers' decisions regarding profile visibility and their participation in discussions about privacy and security. Particularly, it seeks to investigate whether different self-disclosure patterns exist across SO users who involve themselves actively in such discussions, and those who do not.
    %This RQ aims at undertanding zooming into  unveiling 
    \item \textbf{RQ2}: \textit{Are privacy-related constructs associated with users' engagement in cybersecurity discussions?} As with RQ1, former studies have delved into the relation between psychological constructs (e.g., perceived risks and control) and peoples' engagement within OSNs (e.g., \cite{jozani2020privacy,staddon2012privacy}). The purpose of this RQ is to examine whether such correlations also take place in SO but regarding users' participation in discussions about privacy and~security.
    %\item \textit{\textbf{RQ3}: Are there any nuances in profile visibility between SO users with higher and lower levels of engagement?} With this RQ we aim to investigate whether different self-disclosure patterns exist across SO users who actively involve themselves in cybersecurity discussions, and those who do not. Thereby, we seek to determine if profile visibility can be seen as a proxy for targeting developers with personalized content about privacy and security.
\end{itemize}

% %Is users' engagement in cybersecurity discussions associated with any privacy-related construct?

%Engagement can be correlated to a lack of control or high levels of perceived risk, or other constructs

%- Are there any quantitative differences in the profile visibility of SO users with higher and lower engagement levels?

%- Are there any nuances in the profile visibility of SO users with higher and lower engagement levels?

%- Are there any quantitative differences in profile visibility between SO users with higher and lower levels of engagement?

%For this, we analysed the concerns of SO users through (i) their engagement in discussions related to information privacy and security, and (ii) as psychological constructs/antecedents.
To answer these RQs, we have followed a mixed-method approach combining the analysis of data collected from an online survey and information retrieved from SO user profiles. The results of our analysis show significant differences in the self-disclosure practices (i.e., with regard to profile visibility) of users contributing actively to discussions about data protection and information security, and those who do not. These findings not only contribute to a better understanding of users' engagement in such discussions, but also to solutions addressing ``answer-hungry'' questions in Q\&A platforms. Particularly, for the elaboration of incentive strategies and recommender systems promoting the exchange of cybersecurity expertise in SO.

%but also to ongoing research in developer-centred security. Particularly, for the elaboration of support strategies addressing the cybersecurity training of software developers in a personalised way.

\textit{Paper Structure.} Section \ref{sec:related} discusses related work and gives and overview of the paper's theoretical background. Section~\ref{sec:methodology} describes the methodology employed for the study in terms of data collection, aggregation, and survey design. Section~\ref{sec:results} reports the results of our analysis, and Section~\ref{sec:implications} discusses them. Section~\ref{sec:threats} summarises limitations and threats to validity. Section~\ref{sec:conclusions} concludes this work.

\section{Background and Related Work} \label{sec:related}

A growing amount of literature has zoomed into cybersecurity discussions in SO and engagement patterns in OSNs. This section summarises related work elaborating on privacy and security insights gathered through SO. Alongside, we discuss research addressing privacy concerns as a rationale for users' engagement and self-disclosure behaviour in OSNs.

%A growing amount of literature has zoomed into the cybersecurity concerns and practices of software developers. Prior research has aimed at identifying factors that may impair the adoption of privacy- and security-by-design principles, which can ultimately compromise the data-protection standards of information systems. This section summarises related work elaborating on user-centred security, especially through the SO dataset. Alongside, we present and discuss research addressing online self-disclosure behaviour in social network platforms.

%\subsection{Developer-Centred Security} %Privacy and Security Research in SO

\subsection{Cybersecurity Discussions in SO}

Given the Q\&A affordances available within SO, this platform has been widely used as a proxy for understanding the cybersecurity concerns and practices of software engineers \cite{lopez2018sead,Lopez2019,tahaei2020understanding,fischer2017stack}. For instance, \citet{lopez2018sead} conducted a qualitative analysis of SO discussion threads to understand the type of security support developers seek and provide online. Their findings suggest that security-related discussions in SO are rich in terms of technical help but also regarding developers' personal values and attitudes such as trust, fear, and sense of responsibility. In a follow-up article \cite{Lopez2019}, the authors gathered further insights on how security knowledge is built and fostered within the SO community. Overall, their results show that developers often tend towards security-related discussions within the context of technical solutions provided by others. In line with this, \citet{tahaei2020understanding} applied natural language processing techniques to unveil topics emerging within privacy-related questions. The outcome of such an analysis showed that privacy policies, access-control, and encryption are among the main privacy topics addressed by SO members.

%At its core, SO is a peer-production community where knowledge is built from the interaction between developers who ask, comment, correct, argue, and seek to present technical information clearly and concisely \cite{sengupta2020learning}.
At its core, SO is a peer-production community where knowledge is built from the interaction between developers seeking to clarify each other's technical inquires \cite{sengupta2020learning}. Hence, users' participation and engagement are of utmost importance for the sustained development of the platform and the expertise crafted within it. Moreover, timely answers to questions are critical to the platform's efficiency and, thus, to its popularity. Nonetheless, prior research has systematically reported that many questions in SO receive little attention or even remain unanswered/unresolved (up to 30\% by May 2022 \cite{stackExchangesite}). As a catalyst for developers' technical concerns and best practices, it is essential to understand the factors contributing to or impairing users' participation in SO. Prior work has tried to explain why some questions remain unanswered and even proposed machine learning models for predicting whether specific questions will be addressed or not \cite{ahmad2018survey}. Still, the low engagement and the lack of answers to specific questions (including privacy- and security-related ones) remain open issues \cite{gao2020technical}. Hence, there is a call for empirical evidence to (i) help characterise users' engagement in cybersecurity discussions and (ii) elaborate strategies for boosting their participation in such discussions.

\subsection{Insights from Online Social Networks} \label{background_lessons}

Factors influencing people's participation in OSNs have been thoroughly investigated through the lens of privacy concerns. Moreover, prior work has closely analysed users' privacy practices, often accounting for correlations between OSN engagement and self-disclosure behaviour. \citet{staddon2012privacy}, for instance, observed strong associations between privacy concerns and users' engagement on Facebook using an online survey. Their findings revealed that individuals expressing concerns about their privacy also report spending less time on the platform and sharing less content. Hence, they concluded that privacy concerns might play a significant role in people's engagement in OSNs. In line with this, a study by \citet{choi2018instagram} showed that privacy concerns are closely associated with active Instagram use (e.g., sharing content and interacting more with others) and people's selection of a particular OSN platform over others (e.g., Instagram over Snapchat). Alongside, research has systematically reported evidence on the so-called ``privacy paradox'', showing offsets between users' concerns and engagement in OSNs \cite{kramer2019mastering}. Such evidence suggests that, despite expressing privacy concerns, people still join OSNs and disclose significant amounts of personal information.

When it comes to engagement in Q\&A platforms, \citet{kayes2015privacy} investigated the interplay between users' privacy concerns and their participation in Yahoo! Answers. By considering changes in profile visibility as manifestations of privacy concerns, the authors unveiled correlations between users' self-disclosure behaviour and their platform contributions. Overall, they observed that users with a private profile contribute more often and with better content to the platform than those with a public one. Such findings can contribute substantially to the elaboration of Q\&A recommendation approaches. For instance, one could leverage profile visibility for rooting unresolved questions to those users who are more likely to answer them \cite{kayes2015privacy}. Surprisingly, concerns and practices alike have not been thoroughly investigated in SO despite its Q\&A and social network affordances. Moreover, to the extent of our knowledge, the relationship between engagement in cybersecurity topics and self-disclosure practices has not been yet explored nor investigated from a developer-centred perspective.

\section{Methodology}\label{sec:methodology}

We conducted a two-stage empirical study to identify nuances in the self-disclosure practices of users participating actively in cybersecurity discussions, and those who do not. For this, we created a dataset from 7048 SO profiles corresponding to \textit{engaged} and \textit{unengaged} users during the first stage of the study. This dataset was then leveraged on the second stage to conduct an anonymous online survey. Both experimental stages are described in detail in the following subsections.

\subsection{Data Collection} 
To identify users concerned with cybersecurity topics, we first conducted an analysis of privacy- and security-related conversations in SO. Such an analysis consisted in the identification of cybersecurity-relevant conversation threads through their corresponding user-assigned tags. For this, we used SO's Tag Explorer\footnote{See: \url{https://stackoverflow.com/tags}} for the definition of tag sets which were used thereafter to mine relevant conversations. Particularly a set of \textit{topic tags} plus two \textit{language tags} were employed in the identification of cybersecurity-relevant discussions.  

We included \texttt{privacy}, \texttt{privacy-policy}, \texttt{security}, \texttt{code-access-security}, \texttt{data-security}, \texttt{network-security}, and \texttt{gdprconsentform} as topic tags\footnote{It is worth mentioning that, at the moment of conducting this study, these were the only cybersecurity-related tags available in SO.}. Additionally, \texttt{r} and \texttt{python} were used as \textit{language tags} given the increasing popularity of these languages within the data science community \cite{moutidis2021community}. Thereby, we sought to narrow down the scope of the study mainly to data science practitioners as they are prone to handle sensitive data (e.g., medical records, biometric data, demographics). Furthermore, their cybersecurity practices can have a great impact on automated decision-making systems (e.g., biases, discrimination).

\subsubsection{\textbf{Discussions Dataset (D1)}} \label{sec:collection}

\begin{figure*}[t!]
    \centering
    \includegraphics[width=0.8\textwidth]{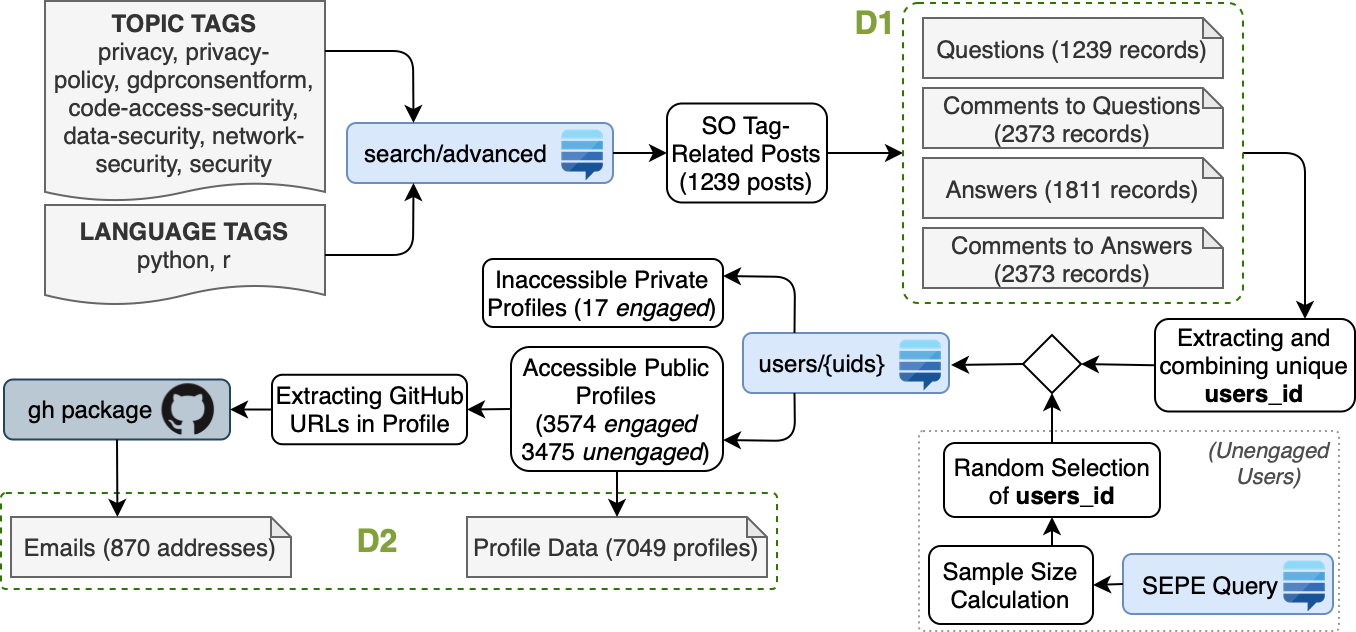}
    \caption{Mining process followed to extract and generate both datasets.}
    \label{fig:miningProcess}
\end{figure*}

Each \textit{topic tag} was explored in combination with each \textit{language tag}, resulting in 14 tag searches. To maximise the size of the dataset, we did not include additional restrictions such as time of posting, the existence of an approved answer, upvotes, or downvotes. Both search and extraction were executed through an R-based mining package included in the StackExchange API\footnote{See: \url{https://api.stackexchange.com/}}. We conducted fourteen independent searches (i.e., one per tag combination) using the \texttt{search/advanced} endpoint and a \texttt{tag filter} provided by the API itself. \textbf{By the end of the mining process, a total of 1239 questions/posts were retrieved from SO (Figure~\ref{fig:miningProcess}).}

Questions posted in SO can be answered or commented on by other platform members. The main difference is that the latter asks for clarification instead of describing a suitable solution. One question can trigger several answers and comments (to the main question or to others' answers) from other SO users interested in the discussion topic. Therefore, such comments and answers are also relevant for identifying SO profiles corresponding to individuals who engage in cybersecurity discussions. Consequently, answers and comments associated with each of the 1239 questions were also mined and included in a \textit{discussions dataset} $D1$. After this additional mining process, \textbf{$D1$ contained 1239 questions, 2558 comments to questions, 1811 answers, and 2373 comments to answers.}

\subsubsection{\textbf{Profiles Dataset (D2)}} \label{dataset_d2}

The information contained in $D_1$ allowed us to identify the SO ids of those users who have either posted a question, provided an answer, or posted a comment deemed as cybersecurity-relevant. Overall, 3591 unique ids were retrieved, from which only 17 corresponded to users with fully private SO profiles. The remaining 3574 ids were used to mine the public information disclosed in their profiles through the StackExchange API (i.e., via the  \texttt{users/\{ids\}} endpoint). 

The email address of some of them was also mined using the GitHub (GH) URL available in the profiles (email addresses are never included in SO profile pages). This step was necessary to recruit participants afterwards for the online survey. This complementary mining process was executed using the R package \texttt{gh}\footnote{See: \url{https://cloud.r-project.org/web/packages/gh/index.html}} resulting in 457 unique e-mail addresses corresponding to \textit{engaged} users. Such information was included in the profiles dataset $D_2$ along with the rest of the profile information extracted from SO.

In order to populate $D_2$ with profile information from \textit{unengaged} users, we first estimated a representative sample size for such a subgroup. For this, we run a query to determine how many users have participated on each language tag\footnote{Query: \url{https://data.stackexchange.com/stackoverflow/query/1392147}} using  Stack Exchange's Data Explorer (SEDE). The result of this query gave 46038 users for the \texttt{r} tag, and 777587 for \texttt{python} tag. Next, we mined the profile information from a representative sample of these two groups with a 99\% confidence and a margin of error of 3\%. Such information was mined directly from the \texttt{users/\{uids\}} endpoint, ensuring that the corresponding SO ids were not already part of the \textit{engaged} group, and were not repeated across each language. \textbf{Overall, we obtained 1830 Python users and 1645 R users (3475 in total)}. These results were merged into the $D_2$ dataset, using an additional variable to indicate whether this information corresponds to \textit{engaged} or \textit{unengaged} users. Like with the \textit{engaged} profiles, we collected the e-mail addresses of 413 \textit{unengaged} users via GH (Figure~\ref{fig:miningProcess}).

\subsection{Data Aggregation}

We parsed the information collected in both datasets to compute two variables of interest: (i) the amount of information users disclose in their profiles, and (ii) their engagement in cybersecurity discussions. The following subsections describe these variables plus an additional analysis we conducted to understand self-disclosure through display names.

\subsubsection{\textbf{Amount of Self-Disclosure}}

SO allows users to include the following information in their profiles: \textit{display name} (with a maximum of 30 characters), \textit{location} (as a text field), \textit{title} (available in the profile, but merged into the display name when using the API), \textit{about me} (HTML-friendly text box of up to 3000 characters), a \textit{website} link, links to \textit{Twitter} and \textit{GitHub} profiles, and a \textit{profile picture} (if not used, the system assigns a randomised avatar). To compute a metric reflecting the amount of personal information revealed in a profile, we assigned a normalised variable (i.e., ranging from 0 to 1) to each field except for the title. The value for each particular variable was estimated as follows:

\begin{itemize}
\item We gave each link (\textit{website}, \textit{Twitter} and \textit{GitHub}) a value of 1 if it was filled in the user's profile, and 0 if not.
\item The \textit{location} variable was calculated as the links (i.e., $1$ if it was completed and $0$ if not). Since users can obfuscate this field (e.g., by using nicknames or aliases), we conducted a card sorting analysis to estimate the reliability of this coding schema. From this analysis, we concluded that location information could be considered accurate if present.
\item The variable corresponding to the \textit{display name} was computed as the proportion of used characters over the total available (30 characters). As with \textit{location}, we completed another card sorting analysis to obtain further reliability insights. Once again, we concluded that the information present in this field could be considered accurate. Both card-sorting analyses can be found in the paper's \textbf{Replication Package}.%\footnote{Available at: \url{https://tinyurl.com/SO-CYBERSEC}}.
\item The \textit{profile image} was retrieved as an URL address during the data collection process. To determine whether an image corresponds to a \textit{custom} or a \textit{default} one we compared its URL against a collection of Gravatar\footnote{See: \url{https://en.gravatar.com/}} URLs (Gravatar pictures are frequently used as default in SO profiles). Using regular expressions, we assigned a 0 value to those profile pictures found in the Gravatar database. Otherwise, they were considered as \textit{custom} and given a value of 1.
\item The \textit{about me} field can have up to 3000 characters allowing HTML formatting. The HTML tags were removed through an R script, and the proportion of used characters was calculated to determine the corresponding disclosure value of this field. This approach assumes that, as more characters are included, more personal information is being revealed.
\end{itemize}

These normalised variables were aggregated into another variable named $soProfDisclosure$ quantifying the amount of personal information disclosed in a SO profile:

\vspace{-2ex}
\begin{equation*}\label{eq:a}
soProfDisclosure = \frac{attsVisibleInProfile}{maxAmountOfDisclosableAtts}
\end{equation*}
\vspace{-2ex}

where $maxAmountOfDisclosableAtts$ corresponds to the maximum number of disclosable attribute values (7 in total), and $attsVisibleInProfile$ to the summation of each normalised variable.

\subsubsection{\textbf{Engagement in Cybersecurity Discussions}}

We classified users into engaged or unengaged, given their participation by computing the number of cybersecurity-relevant questions a user has posted ($\#Q$), the number of answers provided to such questions ($\#A$), and of corresponding comments. This last one was divided into comments to cybersecurity questions ($\#C_Q$) and comments to cybersecurity answers ($\#C_A$). Overall, if the sum $\#Q + \#A + \#C_Q + \#C_A$ was greater than 0, then the user was classified as \textit{engaged} and, otherwise, as \textit{unengaged}.

Also, we classified engaged users into \textit{proactive} and \textit{reactive} according to their tendency towards starting new discussion threads. Particularly, we considered \textit{proactive} users to those who place more questions than comments and answers. That is, in cases where $\#Q \geq \#A + \#C_Q + \#C_A$. Conversely, users posting more comments and answers than cybersecurity questions were classified as \textit{reactive}. That is, when $\#Q < \#A + \#C_Q + \#C_A$.

\subsection{Survey Structure} \label{survey_structure}

To complement the analysis of profile information and discussion threads, we conducted an online survey within a subgroup of SO users. In particular, we aimed at measuring psychological constructs and antecedents to better understand developers' concerns and behaviour regarding cybersecurity. The questionnaire consisted of an introductory part and two main sections:

\begin{enumerate}[i.]
\item The \textbf{introductory section} provided information about the aim of the study along with the conditions for participation/withdrawal (participation was voluntary, and people were given a chance to withdraw at any time). We also included the contact details of the authors in case of further questions and enquiries. 

\item After accepting the survey's terms and conditions, participants were forwarded to the \textbf{first part} of the questionnaire. This part included questions eliciting demographic information (e.g., participants' gender, education level, and current work status) along with their prior experience in software development (e.g., years working with R or Python). 

\item The \textbf{second part} included a set of questions measuring the following constructs: \textit{general privacy concerns} (GPC), \textit{privacy concerns on social threats} (PCS), \textit{privacy concerns on organisational threats} (PCO), \textit{perceived privacy risk} (RSK), \textit{perceived control} (PC), and \textit{self-disclosure} (SD). We used well-established constructs and scales previously elaborated and validated by other authors (i.e., GPC by \citet{Buchanan2007} and the rest by \citet{Krasnova2009}). All questions were close-ended and measured using a 6-Point Likert scale to increase the responses' reliability. We also included an attention question by the end of this section to identify careless respondents and preserve the quality of the results \cite{Kung2018}. %Constructs are summarised in Appendix~\ref{appendix:constructs}. 
\end{enumerate}

This survey was assessed and approved by an Ethics Committee, and is also available in the Replication Package.

% https://drive.google.com/drive/folders/1KoeUYOStTrb9vS8p2wJ3kN9FSUw4yiPk?usp=sharing
% Short: https://tinyurl.com/icse22sec

\subsubsection*{\textbf{Population \& Sampling}}

The survey was distributed through Qualtrics in April/May 2021 using the email addresses collected during the mining process (Section~\ref{dataset_d2}). We gathered 69 responses, out of which five were filtered through the ``attention control'' question. The remaining 64 responses were considered for the corresponding analysis.

\section{Results}\label{sec:results}

We conducted several statistical analyses over the information collected from SO and the responses obtained through the online survey. We conducted a $t$-Test followed by an ANOVA test to identify significant differences in the self-disclosure practices of engaged and unengaged users. The results of these tests were complemented afterwards with an analysis of the survey data.

\subsection{Privacy and Security Discussions (\textit{SO Q\&A data})} \label{sec:qadata}

A total of 1239 cybersecurity-related questions were collected from SO using the StackExchange API (as explained in Section~\ref{sec:collection}). As shown in \autoref{tab:qa_status}, around 67\% of these questions had at least one answer (\textit{answered}), and about 47\% received an answer considered adequate by the user who asked the question (\textit{accepted}). Another 58\% had a positive score (i.e., a positive difference between up-votes and down-votes), whereas 39\% of the questions received at least one comment. SO also allows experienced community members to close questions that are either off-topic or may need further clarification. We observe that around 8\% of the questions in our dataset fall into this category.

%Around 67\% of these questions had at least one answer and about 47\% had an \textit{accepted} answer, where the latter correspond those answers considered adequate by the user asking the question. 

\begin{table}[!ht] %Question status
\caption{Question status indicators.} \label{tab:qa_status}
\centering
%\small
\begin{tabular}{r c c c c c}
\toprule
& \textbf{Has Answers} &  \textbf{Has Accepted Answers}  & \textbf{Has Score $> 0$} & \textbf{Has Comments} & \textbf{Closed}\\
\midrule
Frequency & 825 & 588 & 719 & 489 & 94 \\
\% Total & 67\% & 47\% & 58\% &  39\% & 8\% \\
\bottomrule
\end{tabular}
\end{table}

\subsection{Self-Disclosure Practices (\textit{SO profile data})} \label{results_profile}

\begin{figure}[!h]
    \centering
    \includegraphics[width=0.65\linewidth]{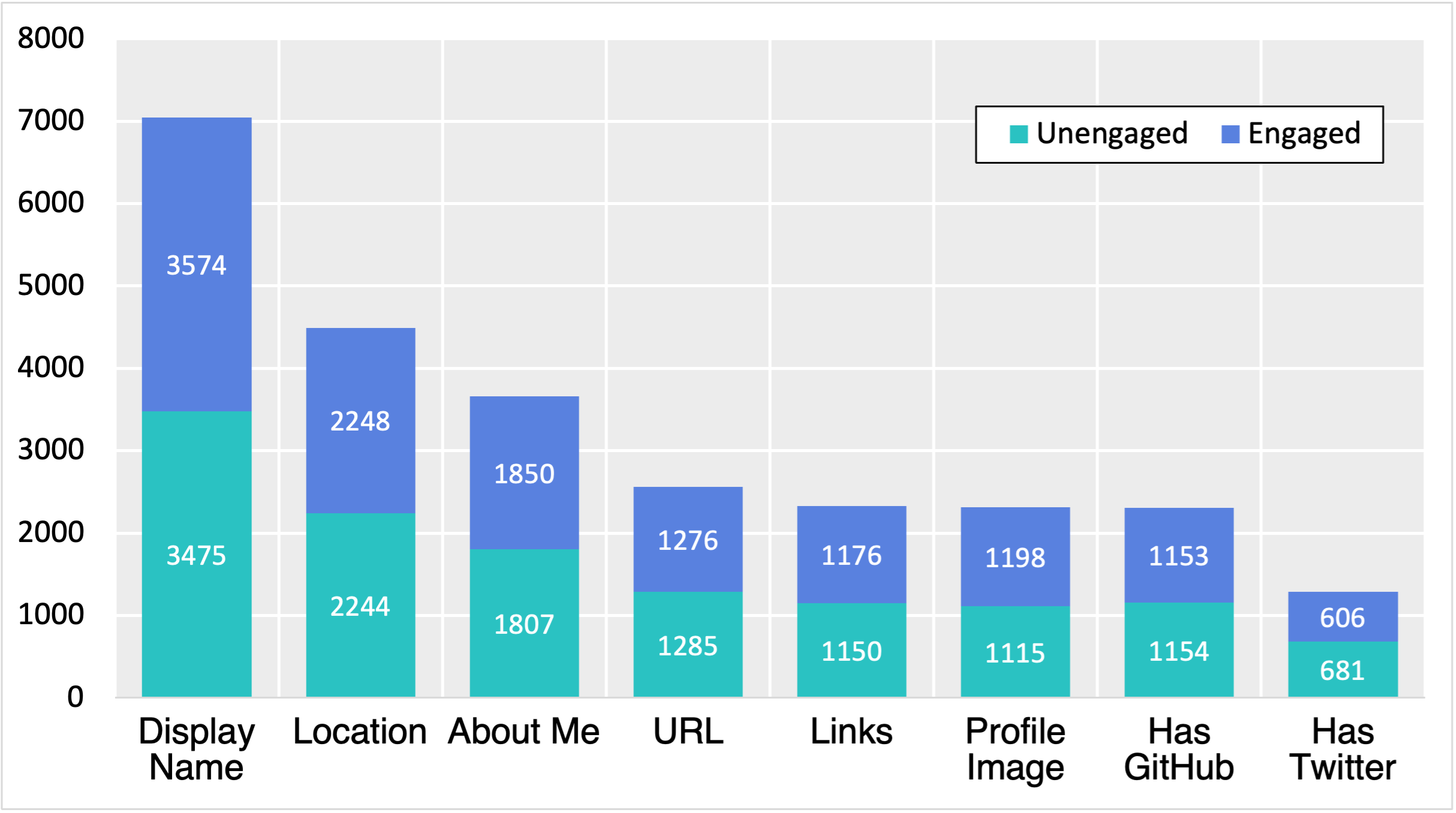}
    \caption{Profile attributes disclosed by \textit{unengaged} and \textit{engaged} users (frequencies).}
    \label{fig:allcats}
\end{figure}

As mentioned in Section~\ref{dataset_d2}, from the 7049 profiles retrieved from SO, 3574 correspond to \textit{engaged} users and 3475 to \textit{unengaged} ones. Figure~\ref{fig:allcats} illustrates the disclosure frequency of each profile attribute in our sample for each group of users. We can see that such frequencies are quite even across all attributes for both groups and that ``display name'' is an attribute everyone discloses. We can also observe that ``location'' and ``about me'' are among the most revealed profile attributes, whereas ``has Twitter'' is the least frequent one.

\begin{figure*}[h!]
    \centering
    \includegraphics[height=4.6cm]{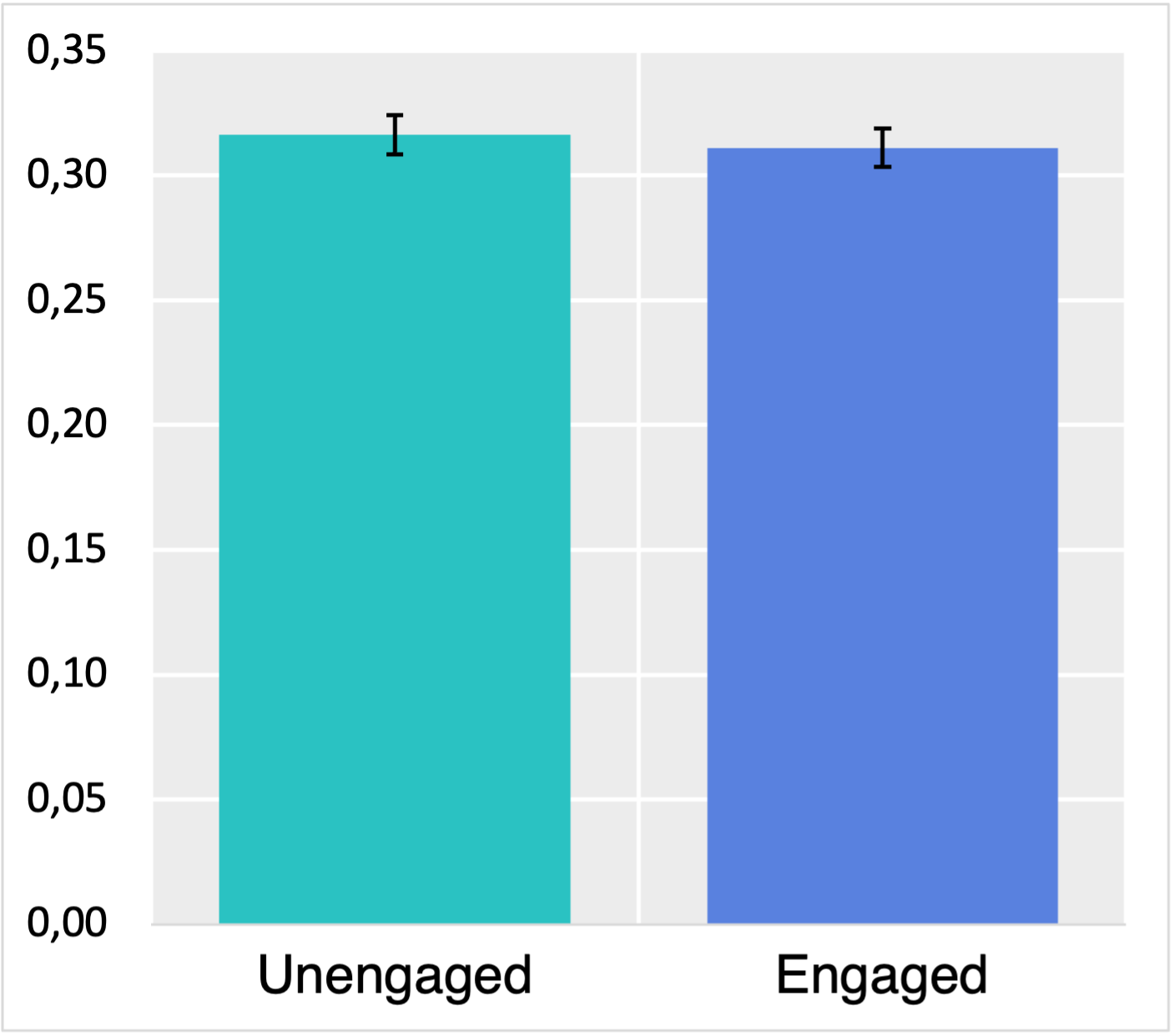}~~
    \includegraphics[height=4.6cm]{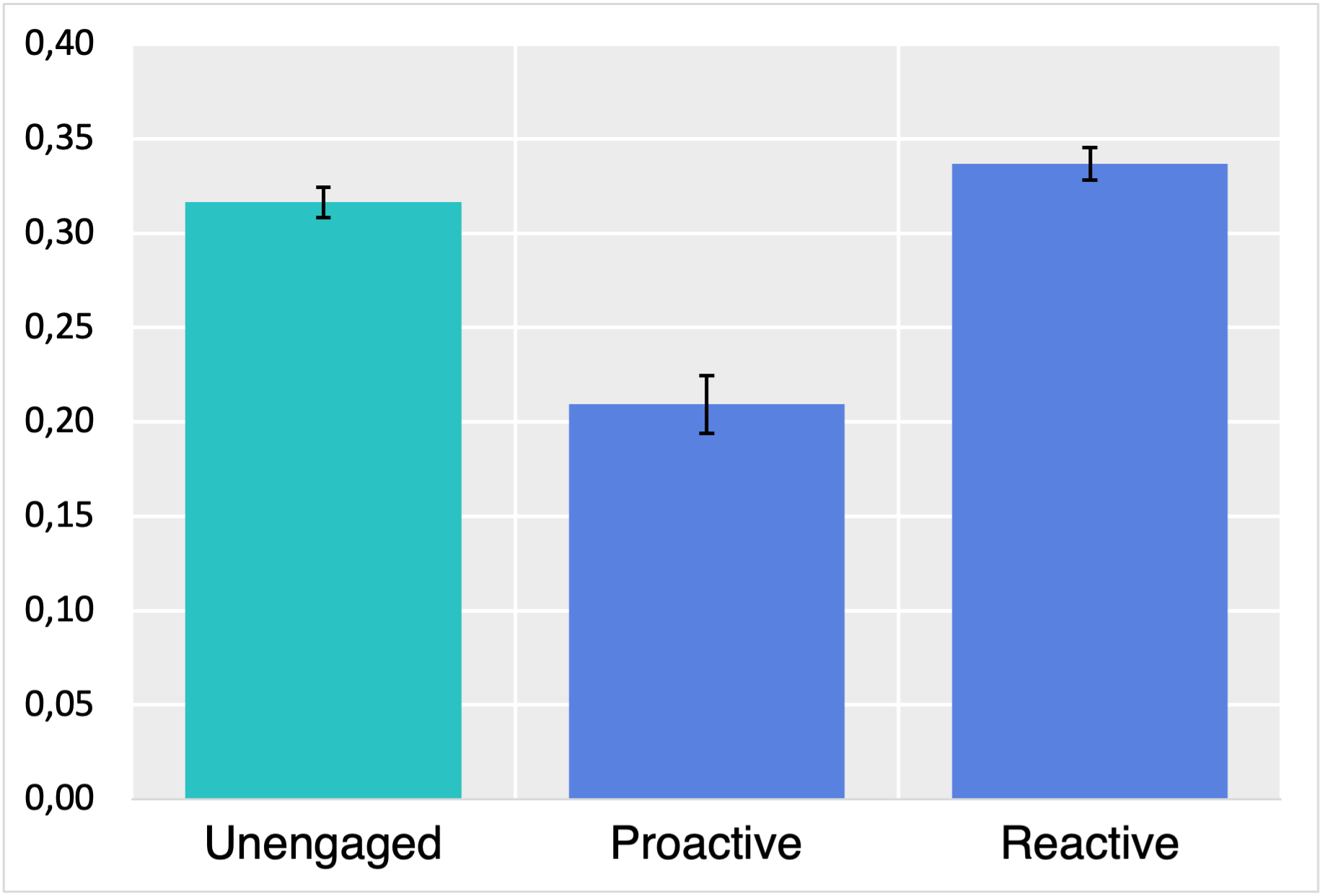}
    \caption{Average self-disclosure of (a) \textit{unengaged} and \textit{engaged} users, and (b) \textit{unengaged}, \textit{proactive}, and \textit{reactive} users.}
    \label{fig:userTypes}
\end{figure*}

We ran an independent samples $t$-Test to identify significant differences in the amount of profile information disclosed by \textit{engaged} and \textit{unengaged} users. Since Levene's test for equality of variances resulted significant ($F_{1,7047}=6.605$, $p=0.10$), the corresponding $t$ statistic was computed without assuming homogeneity of variances. Overall, we found no significant differences in the average amount of self-disclosure between \textit{engaged} and \textit{unengaged} users ($t_{7027.424}=0.918$, $p>0.05$). This can be observed in Figure \ref{fig:userTypes}-a. Hence, we conducted a follow-up ANOVA test to determine whether such differences exist among \textit{unengaged}, \textit{proactive}, and \textit{reactive} users.

% \begin{table}[b!]
%     \caption{One-way ANOVA Test (profile self-disclosure).}
%     \label{tab:1wayanovaData}
% \centering
% \small
% \begin{tabular}{l c c c c c}
% \toprule
% &  \textbf{SS} & \textbf{d.f.} & \textbf{MS} & \textbf{F} & $\boldsymbol{p}$   \\
% \midrule
% Between Groups & 9.340 & 2 & 4.670 & 86.180	& 0.000 \\
% Within Groups & 381.826 & 7046 & 0.054 & & \\	
% Total & 391.167 & 7048 & & & \\		
% \bottomrule
% \end{tabular}
% \end{table}

From the 3574 concerned profiles, 716 corresponded to \textit{reactive} users and 2858 to \textit{proactive} ones. In principle, we can observe differences in the amount of profile information disclosed across these 3 groups (Figure~\ref{fig:userTypes}-b). After conducting the ANOVA test (Table \ref{tab:1wayanovaall}), we could confirm that such differences were indeed statistically significant ($F_{2,7046}=86.180$, $p<0.05$). To determine where these differences actually occur, we ran an additional non-parametric posthoc test. We chose a Games-Howell test since Levene's statistic suggested no equal variances within the sample ($F_{2,7046}=31.772$, $p<0.05$). This analysis revealed significant differences ($p>0.05$) in the average amount of self-disclosure across all paired groups (Table~\ref{tab:anova_posthoc}). That is, between \textit{unengaged-proactive} ($-0.107$), \textit{unengaged-reactive} ($-0.020$), and \textit{proactive-reactive} ($-0.127$).

\begin{table}[!h]
\centering
\begin{threeparttable}

\caption{Games-Howell Test for Differences of Means}
\label{tab:anova_posthoc}

%-----TABLE-----
\begin{tabularx}{0.62\linewidth}{l c c c c}

\toprule
\textbf{Diff. Levels}    & \textbf{Diff. Means}   &   \textbf{SE}          & $\boldsymbol{p}$       &    \textbf{95\% CI}        \\
\midrule

Unengaged\textemdash Proactive    & \phantom{-}0.107*	                & 0.008	        & 0.000	    & (0.086, 0.127)    \\
Unengaged\textemdash Reactive	    & -0.020*	            & 0.006	        & 0.002	    & (-0.034, -0.006)  \\

\midrule

Proactive\textemdash Unengaged	& -0.107*	            & 0.009	        & 0.000	    & (-0.127, -0.086)  \\
Proactive\textemdash Reactive	    & -0.127*	            & 0.008         & 0.000	    & (-0.148, -0.106)  \\

\midrule

Reactive\textemdash Unengaged	    & \phantom{-}0.020*	                & 0.006	        & 0.002	    & (0.006, 0.034)    \\
Reactive\textemdash Proactive	    & \phantom{-}0.127*	                & 0.009	        & 0.000	    & (0.106, 0.148)    \\

\bottomrule
\end{tabularx}
    \begin{tablenotes}
      \small
      \item * The mean difference is significant for $\alpha = 5\%$.
    \end{tablenotes}

\end{threeparttable}
\end{table}

Finally, we conducted a multinomial logistic regression to obtain further insights on the self-disclosure practices of SO users. For this, we considered the unengaged users as the baseline category against which the other groups (i.e., proactive and reactive) should be compared. The parameter estimates of the resulting model are summarised in Table~\ref{tab:mlr_estimates}. As it can be observed, the percentage of information disclosed in a profile (\textit{\% self-disclosure}) is a significant predictor for both proactive and reactive user categories ($p<0.05$). 

On the one hand, for every one-unit increase on \textit{\% self-disclosure}, the likelihood a user has of falling in the proactive category decreases by 2.2\% (i.e., relative to falling in the unengaged group). Conversely, such a likelihood increases by 0.4\% for the reactive category. This model is a significant improvement in fit over an intercept model with no predictors ($\chi^2_{2}=181.388,p<0.001$). However, it does not fit well to the data, which makes it not adequate for prediction purposes (Pearson's $\chi^2_{170}=256.978,p<0.05$).

%-----TABLE-----

\begin{table}[!ht] 
\caption{Multinomial Logistic Regression (estimates).} \label{tab:mlr_estimates}
\centering
%\small
\begin{tabular}{l l c c c c}
\toprule
\textbf{Group} & & \textbf{B} & \textbf{SE} & \textbf{Sig.} & \textbf{Exp(B)}\\
\midrule
proactive & intercept & $-0.991$ & 0.062 & 0.000 & \\	
 & \% self-disclosure & $-0.023$ & 0.002 & 0.000 & 0.978 \\	
reactive & intercept & $-0.313$ & 0.043 & 0.000 & \\
 & \% self-disclosure & $\phantom{-}0.004$ & 0.001 & 0.001 & 1.004 \\	
\bottomrule
\end{tabular}
\end{table}

\subsection{Privacy-Related Constructs (\textit{survey data})}
\label{results_data}

As shown in Table~\ref{tab:demographics}, 76.56\% of the survey respondents worked full time and had more than 10 years of programming experience. Another 75\% reported having more than 5 years of experience working with R or Python, and 56.25\% having a graduate degree. In terms of gender, 61 out of the 64 participants were male, 1 was a woman, 1 non-binary, and 1 preferred not to reveal it.

Following the same user categories investigated in Section~\ref{results_profile}, we conducted a one-way ANOVA test to analyse the privacy-related constructs elicited in the second part of the survey (i.e., GPC, PCS, PCO, RSK, PC, and SD). From the 64 participants, 33 were classified as unengaged, 8 as proactive, and 23 as reactive. Prior to conducting the test, we assessed the reliability of the employed scales by calculating their corresponding Cronbach's Alpha coefficient. In all the cases, such a value was higher than 0.7 suggesting a high internal consistency within each scale's items.

Table~\ref{tab:1wayanovaall} also summarises the outcome of the one-way ANOVA for each constructs measured. We found no significant differences in any of these constructs across proactive, reactive, and unengaged users. This was also the case when conducting a $t$-Test for a two-group classification (i.e., engaged and unengaged).

%-----TABLE-----
\begin{table}[!h]
    \caption{One-way ANOVA Test (profile and survey data).}
    \label{tab:1wayanovaall}

\centering
\begin{tabular}{p{2.5cm} c c c c c}

\toprule
  \makecell[c]{\textbf{Variable}} &   \textbf{SS}    & \textbf{d.f.}  & \textbf{MS}    & \textbf{F}     & $\boldsymbol{p}$   \\
\midrule
\multicolumn{6}{l}{\textit{Profile data}}\\[1ex]
\hspace{1ex}\% self-disclosure & 9.340 & 2 & 4.670 & 86.180 & 0.000 \\
\midrule
\multicolumn{6}{l}{\textit{Survey data}}\\[1ex]
\hspace{1ex}GPC  &	0.825	&  2	& 0.412	& 0.333	&  0.718 \\
\hspace{1ex}PCST &   0.400	& 2	    & 0.200	& 0.141	& 0.869 \\
\hspace{1ex}PCOT &   3.860	& 2	    & 1.930	& 1.127	& 0.331 \\
\hspace{1ex}RSK  &	0.547	& 2	    & 0.274	& 0.384	& 0.682 \\
\hspace{1ex}PC & 2.174	& 2	    & 1.087	& 0.854	& 0.431 \\
\hspace{1ex}SD &	3.082   &	2   & 1.541	& 1.040 & 0.360 \\
\bottomrule

\end{tabular}
\end{table}

\begin{table}[!ht]
\caption{Survey Self-Reported Demographic Data.}
\label{tab:demographics}
\small
\centering

\begin{tabularx}{0.65\linewidth}{lXcc}
\toprule
      \textbf{Demographic}         &      \textbf{Ranges}                         & \textbf{Freq.} &\textbf{ Resp. (\%)}    \\
\midrule

                & Female                        &   1    &   1.56\%      \\
Gender          & Male                          &  61    &  95.31\%      \\
                & Non-Binary                    &   1    &   1.56\%      \\
                & Prefer not to say             &   1    &   1.56\%      \\
                
\midrule

                & Graduate Degree (MSc, PhD)     &  36    &  56.25\%     \\
Educational     & High School or Less           &   3    &   4.69\%     \\
Level           & Some College                  &  11    &  17.19\%     \\
                & Undergrad Degree (BSc, BA)   &  14    &  21.88\%     \\

\midrule

                & Currently in School                         &  1     &   1.56\%     \\
Employment      & Currently in University                     &  5     &   7.81\%     \\
Status          & Unemployed, not looking for work              & 2     & 3.13\% \\
                & Unemployed, looking for work                &  1     &   1.56\%     \\ 
                & Working full-time                           &  49    &  76.56\%     \\ 
                & Working part-time                           &   6    &   9.38\%     \\        

\midrule

Programming     & $<$2 years        & 2     &  3.13\%   \\
Experience      & 2-5 years         & 14    & 21.88\%   \\ 
(R/Python)      & 5-10 years        & 22    & 34.38\%   \\
                & $>$10 years       & 26    & 40.63\%   \\ 
                
\midrule                

Other           & 2-5 years         &  3    &  4.69\%   \\
Programming     & 5-10 years        & 12    & 18.75\%   \\
Experience      & $>$10 years       & 49    & 76.56\%   \\

\bottomrule
\end{tabularx}
\end{table}

\section{Discussion}\label{sec:implications}

This section discusses the results of our study and provides answers to the paper's research questions. We also elaborate on the implications of our findings within the area of developer-centred security, namely the elaboration of strategies for boosting the participation of SO users in cybersecurity discussions.

\subsection{Engagement and Self-Disclosure Behaviour (RQ1)} \label{discussion_rq1}

%\subsubsection*{\textbf{RQ1}: Are users' self-disclosure behaviour associated with their engagement in cybersecurity discussions?} 

Our findings suggest that SO users with a tendency towards starting cybersecurity discussions disclose significantly less information in their profiles than others who do not (Section~\ref{results_profile}). Similar observations were made by \citet{kayes2015privacy} in a study about peoples' engagement in the Q\&A platform Yahoo! Answers. The authors found correlations between users' self-disclosure behaviour (i.e., profile visibility preferences), the frequency, and the quality of their contributions. Particularly, individuals with a more restrictive profile tend to contribute more and with better content than those with a public one. Furthermore, such users also showcase higher retention levels (i.e., average time interval between contributions) and have a higher perception on answer quality.

On the other hand, our results also show that \textit{reactive} users not only reveal more profile information than \textit{proactive} ones, but also more than those \textit{unengaged}. Such a finding is to some extent aligned with prior research on identity formation in Q\&A platforms. To a certain extent, participation in SO is driven by users' need for recognition within the platform. That is, in terms of points and badges that users can assign to each other based on the perceived quality of their contributions \cite{yang2014sparrows}. For instance, a study conducted by \citet{adaji2016towards} showed that high-quality questions are frequently posted by users with complete profile information. \citet{vargo2018identity} also made similar observations and concluded that profile visibility tends to decrease over time. Hence, we could assume that reactive users may also be driven by reputation or recognition when deciding whether to disclose more personal information inside their profiles.

%The authors of \cite{kayes2015privacy} also suggest that users' modifications to their profile visibility settings can be seen as a proxy to their privacy concerns. In such a case, prior research in OSNs has coined results similar to ours. For example, \citet{chen2018revisiting} observed that users with high perceptions of risk and privacy concerns are more prone to restrict the visibility of their profiles than unconcerned ones. Likewise, a survey conducted by \citet{aimeur2019manipulation} also revealed that those who reported having a public profile are more eager to disclose personal information to others in OSNs.

%other authors investigating people's privacy behaviour in OSNs like Facebook. For example, \citet{chen2018revisiting} observed that users with high perceptions of risk and privacy concerns are more prone to restrict the visibility of their profiles than unconcerned ones. Likewise, a survey conducted by \citet{aimeur2019manipulation} also revealed that those who reported having a public profile are more eager to disclose personal information to others in OSNs. Our findings are aligned with these prior ones, particularly if we consider the engagement in cybersecurity discussions as a manifestation of users' privacy and security concerns. 

%\subsubsection*{\textbf{RQ2}: Are privacy-related constructs associated with users' engagement in cybersecurity discussions?} 

\subsection{Engagement and Privacy-Related Constructs (RQ2)} \label{discussion_rq2}

Unlike the results obtained from the users' profile information (Section \ref{results_profile}), the analysis conducted over the survey data showed no significant differences in the elicited constructs (i.e., GPC, PCST, PCOT, RSK, and PC) across \textit{unengaged}, \textit{proactive}, and \textit{reactive} users (Section \ref{results_data}). We hypothesise that this can be related to the relatively good reputation of SO in terms of privacy and data protection, as opposed to OSNs like Facebook. Unlike the latter, SO has not received the attention of mainstream media due to major data-breach scandals or privacy violations. Hence, the role of privacy concerns and perceived risks may not be significant for users' participation and engagement within the platform.

The differences observed in self-disclosure behaviour were not reflected by its survey counterpart (i.e., the SD variable). Nevertheless, and despite that such results may look inconsistent, prior research has also found discrepancies between people's \textit{reported} and \textit{actual} privacy behaviour. As mentioned in Section \ref{background_lessons}, this is often referred to as the ``privacy paradox'', a phenomenon frequently observed within users of OSNs. Our findings suggest traces of this paradox among SO users, especially when contrasting the outcome of the survey analysis with that of the users' profiles. Still, further research is necessary to determine whether the reported privacy behaviour outperforms the actual one across the three user categories. It would be of special interest to understand whether and up to which extent is the privacy paradox manifested among SO users, and how does it relate to their overall engagement.

% \begin{noteBox}
% \textbf{Main findings.}\begin{enumerate}[i.]
% \item \textit{Proactive} users disclose significantly less information in their profiles than \textit{reactive} and \textit{unconcerned} ones.
% \item \textit{Reactive} users reveal on average more personal information in their profiles than \textit{unconcerned} ones.
% \item Traces of \textit{paradoxical behaviour} were observed after contrasting the survey results against the SO profile data.
% \end{enumerate}
% \end{noteBox}
% \vspace{-1ex}

\subsection{Implications and Recommendations} \label{implications_recommendations}

As privacy and security flaws in information systems grow steadily, it is very important to promote the exchange of privacy and security knowledge among software practitioners. To a large extent, the SO community is encompassed by early-career developers seeking for support and guidance in their engineering practices \cite{lopez2018sead}. Hence, it plays a key role in the dissemination and synthesis of cybersecurity expertise. However, our results show an apparent deficit in terms of answers to privacy- and security-related questions (Section~\ref{sec:qadata}). This can not only cause dissatisfaction to those asking such questions, but also damage the platform's value and usefulness in this regard.

%Prior research in developer-centred security has emphasised the importance of tailoring these artefacts to the individual needs of each user in order to increase their efficacy \cite{gorski2020listen,assal2018motivations,thomas2018security}. Under these premises, the results of this work contribute (i) to a better understanding of how developers’ privacy and security concerns are manifested through the Q\&A affordances of SO, and (ii) whether their self-disclosure practices can be used as a proxy for the identification of unconcerned/unsavvy users.

%Having identified nuances in the self-disclosure behaviour across different user groups contributes to the design of personalised training strategies.

\begin{figure*}[h!]
    \centering
    \includegraphics[height=6cm]{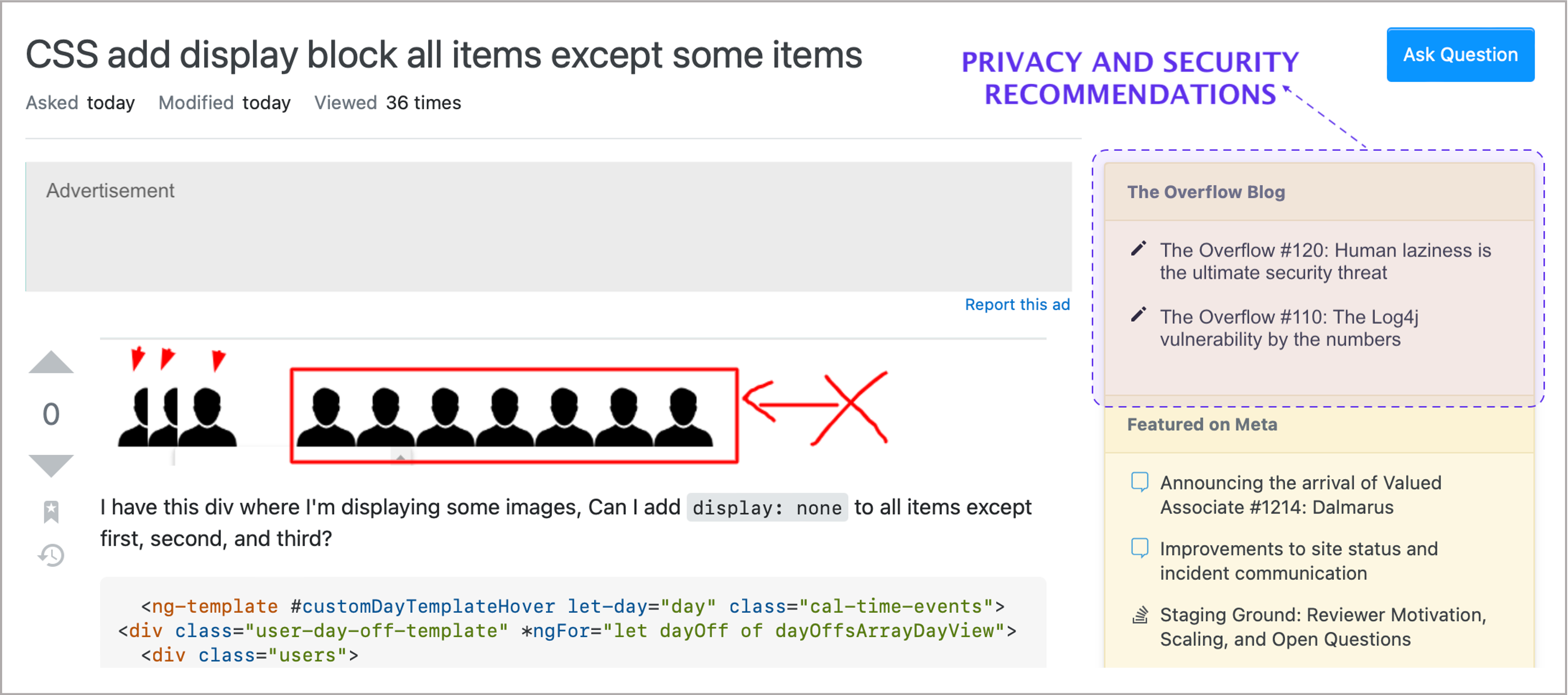}
    \caption{Practical implications (envisaged interface).}
    \label{fig:implications}
\end{figure*}

Having identified nuances in the self-disclosure behaviour across different user groups can be used to foster the exchange of privacy and security expertise. For instance, profile information could be leveraged to motivate the participation in cybersecurity discussions among SO users, by rooting pending questions to those users who are more likely to answer them (e.g., those with a less visible profile). Moreover, closed questions could be assigned to these users for further clarification, and thus increase their resolution chances. Such an approach could also contribute to existing Q\&A recommender systems and frameworks (e.g., \cite{wang2016personalized}) seeking to match forthcoming questions to potential respondents. That is, by incorporating profile visibility as a feature of their question-user matching algorithms.

Similarly, our results could be used to elaborate incentive strategies targeting unengaged individuals. For example, by delivering cybersecurity suggestions to those SO users having a more visible profile. This approach is illustrated in Fig.~\ref{fig:implications}, where a (hypothetically) unengaged user receives such suggestions as she seeks for advice about an issue that is not cybersecurity-related. Here suggestions come in the form of privacy- and security-related entries in the Overflow Blog \cite{overflowBlog2022}, a website curated by SO that gathers essays, opinion articles, and podcasts about computer programming. Using different persuasive styles to approach certain user groups could also improve even more the chances of engagement and behaviour change \cite{aimeur2019manipulation,schaewel2019}. For example, unengaged users could be nudged using a more \textit{authoritarian} style (e.g., \textit{``Microsoft and other big tech companies urge developers to engage in cybersecurity training!''}), whereas a \textit{consensual} one could be applied to proactive and reactive users (e.g., \textit{``Many across the SO community agree: Cybersecurity training is essential for software developers!''}). Likewise, differentiated training content (e.g., access to customised documentation and software artefacts) could be offered to each user group based on a further assessment of their technical skills.

%Finally, having identified scents of paradoxical behaviour among SO users enables future research in developer-centred security. Particularly, regarding its contributing factors and impact over secure software development. Since the privacy paradox has been largely investigated within the scope of OSNs, many findings related to this phenomenon (e.g., the role of peer pressure and subjective norms \cite{barth2017privacy}) could be revised and leveraged for this purpose.

\section{Limitations and Threats to Validity}\label{sec:threats}

To a certain extent, the results of our study are subject to limitations related to its experimental design. In particular, the following \textit{construct}, \textit{external}, and \textit{internal} threats may affect the validity of our findings and conclusions:

\textbf{Construct} threats stem from the degree to which scales, constructs, and instruments measure the properties they intend to \cite{ralph2018construct}. Within the scope of our study, one construct threat arises from the approach employed to compute users' amount of self-disclosure. Profiles are not the only means to reveal private information in SO as users can also disclose personal data inside questions, answers, or comments. However, we conducted our analysis exclusively over SO profiles as they are already adequate and extensive sources of self-disclosure evidence. 

Another construct threat relates to the approach we followed to characterise users' engagement in SO. Indeed, engagement can also take a passive form, where a member (often referred as a ``lurker'') may not contribute actively to a discussion but may still read it and take advantage of its knowledge. We left passive engagement out of the scope of this work as it cannot be determined from the information in our dataset. Still, future research will seek to characterise lurkers and their interaction patterns regarding cybersecurity discussions.

Regarding the psychological constructs elicited during the online survey, we have assessed their reliability by computing the corresponding Cronbach's Alpha coefficient. As mentioned in Section~\ref{results_data}, we obtained values higher than $0.7$ in all the cases, suggesting a high internal consistency of these survey instruments.
%Classification tasks such as the one performed over the display names and locations can also introduce threats to construct validity. For instance, experimenter biases can occur if only one person oversees a card sorting procedure. To mitigate this, the card sorting of display names and locations was conducted by two persons and validated through an agreement discussion (see Replication Package). We obtained a Cohen-Kappa coefficient greater than $+0.8$ in both cases, meaning that our classification approach is highly reliable.

\textbf{External} threats refer to conditions that may affect the generalisability of the study results \cite{cruzes2017threats}. In our case, this relates to the discussions and profile samples we extracted from SO. Since the selection of cybersecurity questions was guided by the tags users assign to them, we may have considered wrongly-tagged questions in our analysis or missed some untagged ones out. Nonetheless, since the SO community of curators often addresses such problems, we assumed the posts we retrieved were accurately labelled.

Another external threat to validity stems from the different sample sizes between Python and R discussions. To minimise this threat, we treated both samples as one without conducting any analysis on each specific language. Likewise, having gathered the email addresses only through GitHub can also be seen as an external threat since it directly impacts the survey's sample size --we sent the survey only to those users from whom we collected their emails via GitHub--. Nevertheless, we managed to gather enough contact details using this approach and distributed the survey to a fair amount of potential respondents.

\textbf{Internal} threats relate to factors that may influence the independent variables of the study in terms of causality~\cite{cruzes2017threats}. In this work, we have analysed the connection between users' self-disclosure practices and their engagement in cybersecurity discussions. However, as mentioned in Section~\ref{discussion_rq1}, both self-disclosure and engagement practices can be influenced by users' need for recognition and popularity within the platform, among other intrinsic and extrinsic factors. Hence, we acknowledge that our study is observational and, as such, cannot be leveraged to draw casual conclusions given the lack of controlled experimental ground truth data.

\section{Conclusions and Future Work}\label{sec:conclusions}

Secure software development largely depends on practitioners' abilities to detect and address potential cybersecurity threats. Still, prior work has shown that many consider security and privacy as secondary aspects of software projects~\cite{acar2016you}. Given the increasing popularity of Q\&A platforms like SO, it is important to characterise and foster the exchange of cybersecurity expertise of their users in order to shape privacy- and security-savvy communities.
%These concerns are frequently out-weighted by other priorities (e.g., functional correctness, time to market, and budget), resulting in flawed data-protection mechanisms and poor security standards.

The results of this work confirm that ``\textit{answer-hungry}'' questions are still a pending issue in SO. Furthermore, it is an issue affecting the privacy- and security-related expertise provided by the platform and its community. As discussed in Section~\ref{implications_recommendations}, having identified different engagement patterns can contribute to elaborating recommender systems and incentive mechanisms targeting this issue. Considering SO's size and outreach, these results could also support the dissemination of privacy- and security-by-design principles among software practitioners. That is, by delivering personalised training programs and tools through the platform to bridge developers' knowledge gaps on cybersecurity. Hence, this work contributes not only to current research in SO but also to ongoing efforts on bringing cybersecurity to the core of software engineering practices. %Hence, this work contributes not only to ongoing research in developer-centred security, but also to current trends in user-centred technologies for behaviour change.

As highlighted in Section~\ref{sec:threats}, the results yielded in this work are observational and call for further investigations. One potential direction for future research is the interplay between privacy concerns and the quality of cybersecurity feedback provided by SO users. For instance, to determine whether developers' \textit{collective privacy concerns} (e.g., their sense of responsibility and empathy towards end-users) and prior cybersecurity experience play a significant role in the extent and frequency of their contributions. For this, we plan to extend our analysis with an empirical study about the factors motivating developers to value and address security and data-protection aspects of their software. For instance, by using scales and survey instruments that capture their efforts towards secure software development, experiences with security issues along with extrinsic motivations and deterrents (similar to the ones proposed in \cite{assal2019think} and \cite{tahaei2021privacy}).

%As highlighted in Section \ref{sec:threats}, several factors may influence people's cybersecurity concerns. So far, we have investigated such concerns in terms of developers' participation in SO discussions, their self-disclosure decisions, and a set of psychological constructs. Nevertheless, the scales we employed during the survey only capture participants' beliefs about certain practices (both personal and organisational) that may negatively affect their \textit{own} privacy boundaries. Consequently, perceptions of data-management practices that may impair the privacy of \textit{others} were not analysed nor considered in this study. \textbf{Future work} will elaborate on developers' \textit{collective privacy concerns} (e.g., their sense of responsibility and empathy towards end-users) as these play a vital role in the implementation of secure software solutions \cite{tahaei2021privacy}. In particular, we will extend our analysis of self-disclosure behaviour in SO with psychological constructs encompassing collective views on privacy and data protection issues (c.f., \cite{morlok2016sharing}).

\section*{Ethical Considerations}

The methodology used in this manuscript was approved by the Australian National University Human Ethics Research Committee (HREC), 
with project code 2021-24127.

\bibliographystyle{ACM-Reference-Format}
\bibliography{references}

%\input{ACM/card_sorting}


\end{document}